          [2001/04/25 1.1 (PWD)]
\documentclass[a4paper,twocolumn]{esapub} 
\usepackage{natbib}

\usepackage{psfig}
\usepackage{epsfig}
\usepackage{graphicx}

\title{XMM-Newton results on the ultracompact 
Low Mass X--ray Binary 4U~1850-087 in the Globular Cluster NGC~6712}
\author{L. Sidoli}
\affil{IASF Milano, Italy}
\author{A.N. Parmar}
\affil{RSSD/ESTEC-ESA Noordwijk, The Netherlands}
\author{T. Oosterbroek}
\affil{INTEGRAL SOC, SODSD, RSSD/ESTEC-ESA Noordwijk, The Netherlands}

\begin{document}


\maketitle

\begin{abstract}
We report on preliminary results from our XMM-Newton observation of the Low Mass X-ray Binary 4U 1850-087, located in the galactic globular cluster NGC 6712. 
It is an ultracompact binary system, with an orbital period of 20.6 min. 
In previous low-resolution X-ray observations 4U 1850-087  displayed a soft excess  residual around 0.7 keV, possibly indicative of the presence of a high Ne/O ratio, as already found in other three ultracompact LMXBs containing a neutron star 
(4U 0614+091, 4U 0918-549, and 4U 1543-624).  
We discuss here our preliminary analysis of high resolution X-ray spectra of 
4U 1850-087, and timing results.
\end{abstract}

\section{Introduction}
 
X1850-087 is an ultracompact Low Mass X-ray Binary (orbital period, P$_{orb}$,
 of 20.6 minutes, Homer et al. 1996)  
located in the globular cluster NGC6712 (at a distance of 6.8 kpc).

Ultracompact X-ray binaries (systems with orbital period less than 80 minutes) 
are believed to contain hydrogen--poor or degenerate companions
(e.g. Verbunt \& van den Heuvel, 1995).

Recent X-ray spectral evidence indicates that these degenerate companions contain anomalously high abundance of neutral Neon 
(e.g. Schulz et al., 2001; Juett et al., 2001).
Indeed, a broad line-like feature at 0.7 keV, 
found in ASCA observations of ultracompact binaries (including X1850-087), 
is suggestive of the presence of an excess in the neutral neon absorption 
local to the systems (Juett et al., 2001). 
High resolution spectroscopy of the ultracompact LMXRBs 2S0918-549 and 4U 1543-624 confirmed an enhanced Ne/O number ratio with respect to the expected ISM ratio Ne/O=0.18 (Juett \& Chakrabarty, 2003).

The aim of our XMM-Newton (Jansen et al. 2001) 
observation is to deeply investigate the presence of Neon excess 
suggested by previous ASCA observation,  
through  high spectral resolution observation with RGS (see den Herder eet al. 2001
for the description of the instrument, covering the energy range 0.3--2 keV).

\section{Observations and Data Analysis}

XMM-Newton data consist of  two observations, 
performed in 2003, September/October, 12 days apart, for a 
net EPIC PN exposure of 8.1ks and 5.9ks, respectively. 
PN operated in Small Window Mode in order to minimize pile-up problems. 
The source rates in the 2 PN observations were $\sim$46~$^{-1}$ and 49~$^{-1}$ 
in the 0.3-12 keV band. 
We will concentrate here only on the RGS spectra. 
The analysis of MOS and PN spectra is still in progress, because
they are particularly complex and cannot be fit with simple models,
because of the presence of structures in the residuals 
at low energies, below 2 keV. 
Indeed, the EPIC PN XMM-Newton spectrum cannot be accounted
for  nor by the ASCA best-fit
(Juett et al. 2001) nor by the BeppoSAX model (Sidoli et al. 2001).

\section{RGS spectral Results}

An accurate visual inspection of the RGS spectra does not
show any evidence for the presence of bright and narrow lines, 
neither in emission nor in absorption (see Fig.~1).

In order to search for neutral Neon excess local to the source,   
we fit the RGS spectrum with a black-body, 
absorbed with a variable-absorption model ({\sc vphabs} in {\sc xspec}) 
and fixing to zero the Neon and Oxygen abundances. 
Then, we included the edges from neutral Neon and Oxygen and after fitting, 
we obtained the Ne and O column densities. 
We got the the following Ne/O number ratio:

\begin{center}
Ne/O = 0.17 $\pm{0.09}$ 
\end{center}

\vskip -0.5cm
to be compared with the ISM value of 0.18. 
Thus we can conclude that there is no evidence for a neutral Neon excess 
local to X1850-087.
\begin{small}
\begin{figure}
\centering
\includegraphics[width=6.0cm,angle=-90]{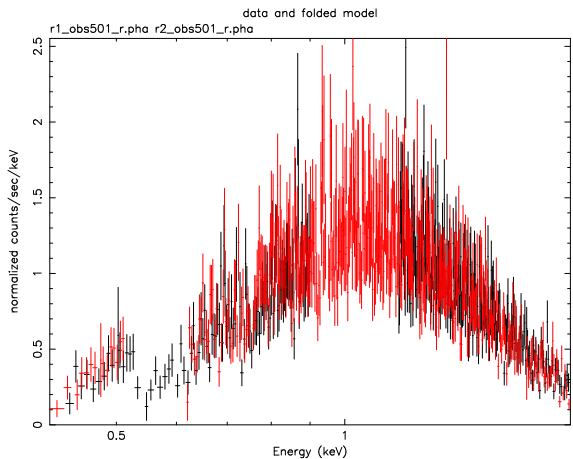}
\caption{RGS1 and RGS2 counts spectra from one of the two XMM-Newton observations. 
There is no evidence for strong and narrow lines, neither in emission 
nor in absorption}
\end{figure}
\end{small}

\section{Preliminary Timing Results: a possible candidate X--ray period at 1258 sec}

A search for periodicity has been performed with epoch folding techniques around the optical period (1233$\pm{4}$~s, Homer et al. 1996).
Our preliminary results shows that 
the  $\chi^2$  distribution reveales a peak at 1258$\pm{20}$~s, 
but only in the second PN observation.
The fact that the X-ray period has been detected only in one of the XMM-Newton observations is puzzling and deserves further investigation.
The double-peaked shape of the folded lightcurve (see Fig.~2)
is possibly indicative of the presence of some symmetric structure in the accretion disk. 
\vskip 0cm
\begin{small}
\begin{figure}[h!]
\centering
\includegraphics[width=4.5cm,angle=-90]{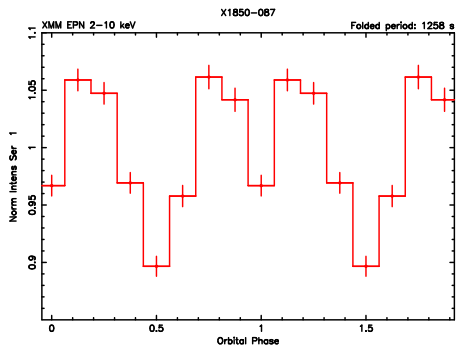}
\caption{ PN   lightcurve folded at the 1258s period (2-10 keV)
}
\end{figure}
\end{small}


\vskip 3cm

\begin{thebibliography}{}

\bibitem[]{}
den Herder, J. W., Brinkman, A. C., Kahn, S. M., et al. 2001, A\&A, 365, L7

\bibitem[]{}
Homer, L., Charles, P. A., Naylor, T., et al., 1996, MNRAS 282, L37

\bibitem[]{}
Jansen, F., Lumb, D., Altieri, B., et al. 2001, A\&A, 365

\bibitem[]{}
Juett, A.M., Psaltis, D., Chakrabarty, D., 2001, ApJ 560, L59

\bibitem[]{}
Juett, A.M., Chakrabarty, D., 2003, ApJ, 599, 498

\bibitem[]{}
Schulz, N.S., Chakrabarty, D., Marshall, H., et al., 2001, ApJ, 563, 941
 

\bibitem[]{}
Sidoli L., Parmar, A.N., Oosterbroek, T., et al., 2001, A\&A 368, 451

\bibitem[]{}
Verbunt F., van den Heuvel E. P. J., 1995, in Lewin W. H. G., van Paradijs J. van den Heuvel E. P. J., eds, X-ray Binaries. Cambridge, Cambridge Univ. Press, p. 457



\end{thebibliography}
\end{document}